\documentclass[12pt]{article}
\usepackage{amssymb,amsmath,amsthm,fullpage,graphicx, bm, bbm, enumerate, setspace, mathtools, float}
\usepackage{natbib}
\usepackage{url}
\usepackage[margin=1in]{geometry}
\usepackage{color}
\usepackage{tikz}
\usepackage{booktabs}
\allowdisplaybreaks

\newtheorem{proposition}{Proposition}[section]

\DeclareMathOperator{\etr}{etr}
\DeclareMathOperator{\Tr}{Tr}

\begin{document}

\title{Monte Carlo simulation on the Stiefel manifold via polar expansion}

\author{Michael Jauch, Peter D. Hoff, and David B. Dunson \\ 
Department of Statistical Science \\ 
Duke University
}
\date{\vspace{-5ex}}
\maketitle

\begin{abstract}
Motivated by applications to Bayesian inference for statistical models with orthogonal matrix parameters, we present $\textit{polar expansion},$ a general approach to Monte Carlo simulation from probability distributions on the Stiefel manifold. To bypass many of the well-established challenges of simulating from the distribution of a random orthogonal matrix $\bm{Q},$ we construct a distribution for an unconstrained random matrix $\bm{X}$ such that $\bm{Q}_X,$ the orthogonal component of the polar decomposition of $\bm{X},$ is equal in distribution to $\bm{Q}.$ The distribution of $\bm{X}$ is amenable to Markov chain Monte Carlo (MCMC) simulation using standard methods, and an approximation to the distribution of $\bm{Q}$ can be recovered from a Markov chain on the unconstrained space. When combined with modern MCMC software, polar expansion allows for routine and flexible posterior inference in models with orthogonal matrix parameters. We find that polar expansion with adaptive Hamiltonian Monte Carlo is an order of magnitude more efficient than competing MCMC approaches in a benchmark protein interaction network application. We also propose a new approach to Bayesian functional principal components analysis which we illustrate in a meteorological time series application. 
\end{abstract}


\section{Introduction}

Probability distributions on the Stiefel manifold, the set of orthogonal matrices \linebreak $\mathcal{V}(k,p) = \left\{\bm{Q} \in \mathbbm{R}^{p\times k} \, \vert \, \bm{Q}^\top \bm{Q} = \bm{I}_k\right\}$ with $p \geq k,$ play a number of roles throughout statistics. The uniform distribution on the Stiefel manifold appears in foundational work on multivariate theory \citep{James1954}, while non-uniform distributions on $\mathcal{V}(k,p)$ arise in modern statistical applications. Distributions on the Stiefel manifold model directions, axes, planes, and rotations in the field of directional statistics \citep{Mardia2009}. They also represent prior or posterior distributions in Bayesian analyses of models with orthogonal matrix parameters. In this work, we are primarily motivated by applications in Bayesian statistics, but the discussion is relevant more broadly.

Statistical models for multivariate data are often naturally parametrized by a set of orthogonal matrices. Parametrization in terms of orthogonal matrices is common in low-rank matrix or tensor estimation, dimension reduction, and covariance modeling. For example, we might model an $n \times p$ data matrix as $\bm{Y}= \bm{U}\bm{D}\bm{V}^\top + \sigma\bm{E},$ where $\bm{U}\in \mathcal{V}(k,n),$ $\bm{V}\in\mathcal{V}(k,p),$ $\bm{D}$ is a $k\times k$ diagonal matrix with positive entries on the diagonal, $\bm{E}$ is a matrix of errors, and $\sigma > 0.$ This model and variants are important in matrix denoising problems \citep{Donoho2014} and model-based principal component analysis (PCA) \citep{Hoff2009}.

Bayesian analyses of models with orthogonal matrix parameters are increasingly common but raise computational challenges. In modern Bayesian statistics, analytic calculation of posterior expectations or exact Monte Carlo simulation from the posterior is typically infeasible. Instead, one constructs a Markov chain whose stationary distribution is the posterior using Markov chain Monte Carlo (MCMC) methods. For models with orthogonal matrix parameters, this Markov chain must lie on the Stiefel manifold. However, the constraints which define the manifold complicate MCMC simulation to the extent that Bayesian analyses of models with orthogonal matrix parameters are often prohibitively difficult.

A number of authors have addressed simulation from distributions on the Stiefel manifold, but there remains a need for more routine and flexible methodology for posterior simulation in models with orthogonal matrix parameters. In the directional statistics literature, which focuses on exact Monte Carlo simulation in a low dimensional setting, rejection sampling is common. See, for example, \citet{Kent2013}. These rejection sampling approaches are not well suited for routine and flexible posterior simulation, as they must be tailored to particular distributions, and acceptance rates can decrease rapidly with increasing dimension or concentration of the target distribution. \citet{Hoff2009} proposes a Gibbs sampler for the Bingham-von Mises-Fisher family of distributions on the Stiefel manifold and applies it to posterior simulation for the network eigenmodel discussed in Section \ref{eigenmodel}. The Gibbs sampler of \citet{Hoff2009} can be a practical option for posterior simulation but is applicable only when the conditional posterior distributions belong to the designated family. As we will see in Section \ref{eigenmodel}, Gibbs sampling can also produce Markov chains with high autocorrelation. Furthermore, simulation from conditional distributions is performed via rejection sampling, and acceptance rates can be vanishingly small, as described in \citet{Brubaker2012}. \citet{Byrne2013} introduce geodesic Monte Carlo (GMC), an elegant and well-motivated algorithm extending Hamiltonian Monte Carlo (HMC) \citep{Neal2011} to distributions defined on the Stiefel manifold and other manifolds embedded in Euclidean spaces. However, without methodology for adaptive tuning parameter selection or a robust software implementation, GMC does not yet offer routine and flexible posterior simulation. \citet{Jauch2018} and \citet{Pourzanjani2017} reparametrize the Stiefel manifold in terms of unconstrained Euclidean parameters, derive the Jacobian term required to map the target distribution from the Stiefel manifold to Euclidean space, then leverage MCMC software to simulate from the transformed distribution. The core idea of recasting a constrained simulation problem as an easier unconstrained problem is compelling, but the cost of computing the Jacobian term (in \citet{Jauch2018}) and the pathologies introduced in mapping between topologically distinct spaces are drawbacks of these reparametrization approaches. 

In this work, we present $\textit{polar expansion},$ a general approach to Monte Carlo simulation from probability distributions on the Stiefel manifold. To bypass many of the well-established challenges of simulating from the distribution of a random orthogonal matrix $\bm{Q} \in \mathcal{V}(k,p),$ we construct a distribution for an unconstrained random matrix $\bm{X}\in \mathbbm{R}^{p\times k}$ such that $\bm{Q}_X,$ the orthogonal component of the polar decomposition, is equal in distribution to $\bm{Q}.$ The distribution of $\bm{X}$ is amenable to Markov chain Monte Carlo simulation using standard methods, and an approximation to the distribution of $\bm{Q}$ can be recovered from a Markov chain on the unconstrained space. When combined with modern MCMC software, polar expansion allows for routine and flexible posterior inference in models with orthogonal matrix parameters. Polar expansion can be seen as a generalization of the method for simulating from the unit sphere $\mathcal{V}(1,p)$ built into Stan at the time of writing \citep{StanDevelopmentTeam2019}.

We provide an outline of what follows. In Section \ref{polar_expansion}, we present polar expansion in detail. In Section \ref{exactMC}, we build intuition through simple examples in which exact Monte Carlo simulation is possible. That discussion serves as a prelude for Section \ref{MCMC}, which addresses polar expansion and MCMC simulation in more complex settings, including posterior simulation for models with orthogonal matrix parameters. In Section \ref{applications}, we illustrate the practical importance of polar expansion in applications. We find that polar expansion with adaptive HMC is an order of magnitude more efficient than competing MCMC approaches in a benchmark protein interaction network application. We also propose a new approach to Bayesian functional principal components analysis which we illustrate in a meteorological time series application. We conclude with a brief discussion in Section \ref{discussion}. Code to reproduce the figures and analyses in this article is available at \url{https://github.com/michaeljauch/polar}.

\section{Polar expansion via change of variables} \label{polar_expansion}

%Polar expansion is a general approach to Monte Carlo simulation from distributions on the Stiefel manifold. As we described in the introduction, simulating from the distribution of a random orthogonal matrix $\bm{Q} \in \mathcal{V}(k,p)$ is fraught with challenges, motivating attempts to transform this constrained simulation problem into an easier unconstrained problem. With polar expansion, we recast the problem by constructing an unconstrained random matrix $\bm{X}\in \mathbbm{R}^{p\times k}$ for which $\bm{Q}_X \stackrel{d}{=} \bm{Q}.$ With the arrival of MCMC software such as Stan, simulating from the distribution of $\bm{X}$ is straightforward compared to the original constrained problem. In this subsection, we present the construction of $\bm{X}$ in detail. 

The polar decomposition is the unique representation of a full rank matrix $\bm{X} \in \mathbbm{R}^{p\times k}$ as the product $\bm{X} = \bm{Q}_X \bm{S}_X^{1/2}$ where $\bm{Q}_X \in \mathcal{V}(k,p),$ $\bm{S}_X$ is a $k \times k$ symmetric positive definite (SPD) matrix, and $\bm{S}_X^{1/2}$ is the symmetric square root of $\bm{S}_X.$ As the name suggests, the polar decomposition is analogous to the polar form $z=e^{i\varphi}r$ of a nonzero complex number, with $\bm{Q}_X$ being the analog of $e^{i\varphi}$ and $\bm{S}_X^{1/2}$ being the analog of $r.$ The components of the polar decomposition can be computed from $\bm{X}$ as $\bm{Q}_{X} = \bm{X}(\bm{X}^\top \bm{X})^{-1/2}$ and $\bm{S}_{X}=\bm{X}^\top\bm{X}.$ In terms of the singular value decomposition $\bm{X}=\bm{U}\bm{D}\bm{V}^\top,$ we have $\bm{Q}_X=\bm{U}\bm{V}^\top$ and $\bm{S}_X^{1/2}=\bm{V}\bm{D}\bm{V}^\top.$ Additionally, the orthogonal component $\bm{Q}_{X}$ has an intuitive geometric interpretation as the closest matrix in $\mathcal{V}(k,p)$ to $\bm{X}$ in the Frobenius norm, i.e. $\bm{Q}_{X} = \text{argmin}_{\bm{Q}\in \mathcal{V}(k,p)} \|\bm{X} - \bm{Q}\|_{F}.$

Given a density $f_Q$ defined with respect to the uniform measure on $\mathcal{V}(k,p),$ we would like to simulate a random orthogonal matrix $\bm{Q}$ whose distribution has density $f_Q.$ Our strategy, motivated by the relative ease of unconstrained simulation, is to simulate a random matrix $\bm{X}$ from a distribution whose $\bm{Q}_X$-margin has density $f_Q$. The distributions on $\bm{X}$ that have the desired marginal distribution for $\bm{Q}_X$ can be identified via a change of variables. The mapping from a real, full rank matrix $\bm{X}$ to the components $(\bm{Q}_X, \bm{S}_X)$ of its polar decomposition is one-to-one, so the density of the distribution of $\bm{X}$ can be derived from the density of the joint distribution of $\bm{Q}_X$ and $\bm{S}_X$ as 
\begin{align*}
  f_X(\bm{X}) = f_{S_X|Q_X}(\bm{S}_X \mid \bm{Q}_X)\, f_{Q_X}(\bm{Q}_X) \times J(\bm{Q}_X, \bm{S}_X; \bm{X}). 
\end{align*} 
The Jacobian of the transformation from $\bm{X}$ to $(\bm{Q}_X, \bm{S}_X)$ is provided in \citet{Chikuse2003}: 
\begin{align*}
    J\left(\bm{Q}_X, \bm{S}_X ; \bm{X}\right) &=   \frac{\Gamma_k(\frac{p}{2})}{\pi^\frac{pk}{2}}\left|\bm{S}_X\right|^{-\frac{p-k-1}{2}}.
\end{align*}
If $\bm{Q}_X$ is to have marginal density $f_Q$, we must have 
\begin{align*}
  f_X(\bm{X}) = f_{S_X|Q_X}(\bm{S}_X|\bm{Q}_X)\,  f_Q(\bm{Q}_X) \times   J(\bm{Q}_X, \bm{S}_X; \bm{X}).
\end{align*} 
Putting these observations together, we arrive at the following proposition: 
\begin{proposition} \label{X_density_prop}
The $\bm{Q}_X$-margin of an absolutely continuous random matrix $\bm{X}$ has density $f_Q$ if and only if 
\begin{align}
 f_X(\bm{X}) = f_{S_X|Q_X}(\bm{S}_X|\bm{Q}_X)\,  f_Q(\bm{Q}_X) \times   J(\bm{Q}_X, \bm{S}_X; \bm{X}). \label{X_density_eqn}
\end{align}
\end{proposition}

There is not a unique distribution for $\bm{X}$ which has the desired $\bm{Q}_X$-margin. From Proposition \ref{X_density_prop}, we see there is one such distribution for each choice of conditional density $f_{S_X|Q_X}.$ For some simulation problems, there is an obvious choice for the distribution of $\bm{X}$ having the desired $\bm{Q}_X-$margin, and the conditional density $f_{S_X|Q_X}$ is an afterthought. For others, there is no obvious choice. In that case, we construct a distribution for $\bm{X}$ by choosing a conditional density $f_{S_X|Q_X}$ and plugging it into Equation (\ref{X_density_eqn}).  

The term ``parameter expansion" applies to methods which expand the parameter space of a statistical model by introducing redundant working parameters for computational purposes. The working parameters render the expanded parametrization non-identifiable, but the original parameters of interest can still be recovered. Parameter expansion has been successfully applied in the context of the expectation maximization algorithm \citep{Liu1998} and MCMC simulation \citep{Liu1999, VanDyk2001}. As the name suggests, polar expansion fits this pattern. When applied to posterior simulation in a model with a parameter $\bm{Q}\in \mathcal{V}(k,p),$ polar expansion replaces the orthogonal matrix $\bm{Q}$ having $pk - k(k-1)/2$ free parameters with an unconstrained matrix $\bm{X}\in \mathbbm{R}^{p\times k}$ having $pk$ free parameters. The expanded model is non-identifiable, but the original parameter of interest $\bm{Q}$ can be recovered via the polar decomposition of $\bm{X}.$

\section{Polar expansion and exact Monte Carlo} \label{exactMC}

There are some simple, well-known distributions for a random orthogonal matrix $\bm{Q}$ which are the $\bm{Q}_X$-margin of a standard distribution for $\bm{X}.$ If exact Monte Carlo simulation of $\bm{X}$ is possible, then the same is true of $\bm{Q}.$ To simulate a random orthogonal matrix $\bm{Q}$ with the desired distribution, we simply simulate $\bm{X}$ and then set $\bm{Q}=\bm{Q}_X.$ We go through the following examples, in order of increasing generality, to build intuition about polar expansion and familiarity with the required calculations. We will draw on these foundations in Section \ref{MCMC}, which addresses polar expansion in more complex settings. 

\paragraph{Uniform distribution on the sphere}
Suppose we want to simulate a random vector $\bm{Q}$ which is uniformly distributed on the unit sphere $\mathcal{V}(1,p).$ A well-known approach described, for example, in  \citet{Marsaglia1972} is to simulate $\bm{X} \sim N(\bm{0}, \bm{I}_p)$ and then set $\bm{Q} = \bm{Q}_X = \bm{X}/\sqrt{\bm{X}^\top\bm{X}}.$ The random variable $S_X > 0$ is independent of $\bm{Q}_X$ and $\chi^2_p$ distributed.

\paragraph{Uniform distribution on the Stiefel manifold}
Now suppose we want to simulate a random orthogonal matrix $\bm{Q}$ which is uniformly distributed on the Stiefel manifold $\mathcal{V}(k,p).$ We can do so by simulating a random matrix $\bm{X} \in \mathbbm{R}^{p\times k}$ with independent standard normal entries and then setting $\bm{Q} = \bm{Q}_{X}.$ This construction of a uniform orthogonal matrix is also well-known \citep{Eaton1989}. The random SPD matrix $\bm{S}_X$ is independent of $\bm{Q}_X$ and Wishart $\text{W}_p(\bm{I}_k)$ distributed.

\paragraph{Matrix angular central Gaussian}
The random orthogonal matrix $\bm{Q}$ is said to have a matrix angular central Gaussian $\text{MACG}(\bm{\Sigma})$ distribution if $\bm{Q} \stackrel{d}{=} \bm{Q}_X$ where $\bm{X}\sim N_{p,k}(\bm{0},\bm{\Sigma}, \bm{I})$ \citep{Chikuse2003}. The notation $N_{p,k}(\bm{0},\bm{\Sigma}, \bm{I})$ indicates a centered matrix normal distribution  with $\bm{\Sigma}$ as its row covariance matrix and the identity as its column covariance matrix \citep{Srivastava1979, Dawid1981}. The $\text{MACG}(\bm{\Sigma})$ distribution has density $f_{\bm{Q}}(\bm{Q}) =|\bm{\Sigma}|^{-k/2}|\bm{Q}^\top\bm{\Sigma}^{-1}\bm{Q}|^{-p/2}$ and is uniform on the Stiefel manifold when $\bm{\Sigma} = \bm{I}.$ Clearly, we can simulate $\bm{Q}\sim \text{MACG}(\bm{\Sigma})$ by first simulating $\bm{X} \sim N_{p,k}(\bm{0},\bm{\Sigma}, \bm{I})$ and then setting $\bm{Q} = \bm{Q}_X.$ The random SPD matrix $\bm{S}_X$ is independent of $\bm{Q}_X$ with
\begin{align} 
f_{S_X \mid Q_X}(\bm{S}_X \mid \bm{Q}_X) =\frac{ {}_0F_0^{(p)}\left(-\frac12 \bm{\Sigma}^{-1}, \bm{S}_X \right)}{2^{pk/2}\Gamma_k(\frac{p}{2})\left|\bm{\Sigma}\right|^{k/2}} \left|\bm{S}_X\right|^{(p-k-1)/2}. \label{SMACG}
\end{align} See \citet{Chikuse2003} for a discussion of the hypergeometric function ${}_0F_0^{(p)}$ of matrix argument.

As we indicated before, the examples are listed in order of increasing generality. In each case, the distribution of $\bm{Q}$ is MACG. More generally, any distribution which is the $\bm{Q}_X$-margin of a standard distribution for $\bm{X}$ lends itself to exact Monte Carlo simulation via polar expansion.

\section{Polar expansion and MCMC} \label{MCMC}

In many simulation problems of interest, the target distribution of the random orthogonal matrix $\bm{Q}$ is not the $\bm{Q}_X$-margin of a standard distribution for $\bm{X}.$ While exact Monte Carlo simulation from these distributions is out of reach, we can still apply polar expansion to construct a distribution for $\bm{X}$ which has the desired $\bm{Q}_X$-margin and is amenable to MCMC simulation. We first consider the scenario in which the distribution of $\bm{Q}$ is a posterior arising from an MACG prior. Guided by the examples of the previous section, we propose a simple way to construct a distribution for $\bm{X}$ with the desired $\bm{Q}_X$-margin. We then consider the very general scenario in which the distribution of $\bm{Q}$ is specified by a density $f_Q$ which is known up to a multiplicative constant. In this general scenario, we construct a distribution for $\bm{X}$ which has the desired $\bm{Q}_X$-margin by choosing a conditional density $f_{S_X \mid Q_X}$ and plugging it into Equation (\ref{X_density_eqn}). Finally, we motivate our recommendation of HMC for MCMC simulation from the distribution of $\bm{X}.$

\subsection{Posterior simulation with an MACG prior}

We consider the case in which the distribution of $\bm{Q}$ is a posterior arising from an $\text{MACG}(\bm{\Sigma})$ prior. The $\text{MACG}(\bm{\Sigma})$ distribution is uniform when $\bm{\Sigma} = \bm{I}$ but can incorporate prior structure such as row dependence when $\bm{\Sigma}\neq \bm{I}.$ We take advantage of this flexibility in the functional PCA application of Section \ref{functionalPCA}.

Suppose we have data $\bm{y}$ whose distribution given the unknown parameter $\bm{Q} \in \mathcal{V}(k,p)$ has density $p(\bm{y} \, \vert \, \bm{Q}).$ The MACG prior density is $p(\bm{Q})=|\bm{\Sigma}|^{-k/2}|\bm{Q}^\top\bm{\Sigma}^{-1}\bm{Q}|^{-p/2}$ and the posterior density satisfies $p(\bm{Q}\, \vert \, \bm{y}) \propto p(\bm{y} \, \vert \, \bm{Q}) \, p(\bm{Q}).$ To approximate the posterior distribution of $\bm{Q},$ we propose constructing a Markov chain  $\{\bm{X}_t\}_{t=1}^{T}$ whose stationary distribution has density
\begin{align}
    f_{X}(\bm{X}) &= p(\bm{X} \mid \bm{y}) \nonumber \\
    &\propto p(\bm{y} \mid \bm{Q}_X) \, N_{p,k}(\bm{X} \mid \bm{0}, \bm{\Sigma}, \bm{I}) \label{MACGposterior}
\end{align} 
and taking $\{\bm{Q}_{X_t}\}_{t=1}^{T}$ as our approximation. The $\bm{Q}_X$-margin of the distribution for $\bm{X}$ specified by the density (\ref{MACGposterior}) is the posterior distribution of $\bm{Q}.$ This can be verified formally via a change of variables from $\bm{X}$ to the components of its polar decomposition. The distribution of $\bm{X}$ is nonstandard, but knowing its density allows us to apply standard MCMC methods.

An analogous approach to posterior simulation is available whenever the prior distribution for $\bm{Q}$ is the $\bm{Q}_X$-margin of a standard distribution for $\bm{X}.$ One can simply replace the matrix normal density in Equation (\ref{MACGposterior}) with the alternative density for $\bm{X}.$  We emphasize the MACG distribution because of its utility as a prior distribution and because, as far as we are aware, it is the only distribution in the literature which is the $\bm{Q}_X$-margin of a standard distribution for $\bm{X}.$

\subsection{General simulation problems}

There are important settings in which the distribution of an orthogonal matrix $\bm{Q}$ is neither the $\bm{Q}_X$-margin of a standard distribution for $\bm{X}$ nor a posterior arising from such a prior. In particular, the distribution of $\bm{Q}$ might belong to the Bingham-von Mises-Fisher family \citep{Hoff2009} or be a posterior distribution arising from a prior which is not the $\bm{Q}_X$-margin of a standard distribution for $\bm{X}.$ With these examples in mind, we consider simulating from a distribution for $\bm{Q}$ specified by a density $f_Q$ which is known up to a multiplicative constant. 

In this general scenario, unlike the previous examples, there is no obvious choice for the distribution of $\bm{X}$ which has the desired $\bm{Q}_X$-margin. Instead, we construct a distribution for $\bm{X}$ by choosing a conditional density $f_{S_X \mid Q_X}$ and plugging it into Equation (\ref{X_density_eqn}). We propose to let $f_{S_X \mid Q_X} = \text{W}_p(\bm{S}_X; \bm{I}_k).$ That is, the conditional density $f_{S_X \mid Q_X}$ is a Wishart density with $p$ degrees of freedom and $\bm{I}_k$ as its scale matrix. With this choice, the density of the distribution of $\bm{X}$ simplifies to
    \begin{align} \label{Xdensity}
    f_{X}(\bm{X}) &= (2\pi)^{-pk/2} \etr\left(- \bm{X}^\top\bm{X}/2\right) f_{\bm{Q}}(\bm{Q}_X).
    \end{align}
When $f_{\bm{Q}}(\bm{Q}_X) \propto 1$ and the distribution of $\bm{Q}$ is uniform, the entries of $\bm{X}$ are independent standard normal random variables. This appealing correspondence between the uniform distribution on $\mathcal{V}(k,p)$ and the distribution of $pk$ independent standard normals is one motivation for our choice of conditional density $f_{S_X \mid Q_X}.$ Furthermore, when applied to the problem of simulating from the unit sphere $\mathcal{V}(1,p),$ our proposed approach is equivalent to the method for simulating from $\mathcal{V}(1,p)$ built into Stan at the time of writing \citep{StanDevelopmentTeam2019}.

\subsection{Hamiltonian Monte Carlo}

To simulate from the distribution of $\bm{X},$ we recommend Hamiltonian Monte Carlo \citep{Neal2011}. Hamiltonian Monte Carlo (originally Hybrid Monte Carlo \citep{Duane1987}) is a class of MCMC methods which simulates Hamiltonian dynamics in order to propose long distance moves in the state space while maintaining high acceptance rates. Markov chains produced by HMC typically converge more quickly to their stationary distribution and exhibit less autocorrelation than those produced by random walk Metropolis or Gibbs sampling algorithms. Through their automatic differentiation and adaptive tuning functionality, software implementations such as Stan \citep{Carpenter2017} greatly simplify applications of HMC. They also provide a powerful set of diagnostics which alert the user to potential problems that may lead to poor Monte Carlo estimates.

\section{Applications} \label{applications}
\subsection{Network eigenmodel for protein interaction data} \label{eigenmodel}

We compare polar expansion to competing MCMC approaches in a benchmark protein interaction network application. Using polar expansion with adaptive HMC as implemented in Stan, GMC without parallel tempering, and the Gibbs sampler of \citet{Hoff2009}, we simulate from the posterior distribution of the network eigenmodel of \citet{Hoff2009} applied to the protein interaction data first appearing in \citet{Butland2005}. Compared to GMC, its strongest competitor, polar expansion with adaptive HMC is an order of magnitude more efficient in terms of effective sample size per iteration and comparable in terms of iterations per second. 

The application which we use as a benchmark was first introduced in \citet{Hoff2009}. The interactions of $p=270$ proteins of \textit{Escherichia coli} are recorded in the binary, symmetric $p\times p$ matrix $\bm{Y} = (y_{i,j}).$ If protein $i$ and protein $j$ interact, then $y_{i,j}=1.$ Otherwise, $y_{i,j}=0.$ The edge probabilities are assumed to have a low-rank structure with 
\begin{align}
P(y_{i,j} = 1) = \Phi\left[c + \left(\bm{Q}\bm{\Lambda}\bm{Q}^\top\right)_{i,j}\right] %\label{network_prob}
\end{align}
where $\Phi$ is the cumulative distribution function of a standard normal random variable and $(c, \bm{Q}, \bm{\Lambda})$ are unknown parameters. The parameter $\bm{Q}$ is a $p \times 3$ orthogonal matrix, $\bm{\Lambda}=\text{diag}(\lambda_1, \lambda_2, \lambda_3)$ is a $3 \times 3$ diagonal matrix, and $c$ is a real number. Following \citet{Hoff2009}, $\bm{Q}$ is \textit{a priori} uniform on $\mathcal{V}(3,p),$ the diagonal elements of $\bm{\Lambda}$ have independent $N(0,p)$ prior distributions, and $c \sim N(0,10^2).$ 

\citet{Hoff2009} proposes a Gibbs sampler for posterior simulation. As discussed in \citet{Albert1993}, the probit link function admits a simple data augmentation scheme which often leads to standard conditional posterior distributions. After taking advantage of this data augmentation scheme, the conditional posterior distribution of the orthogonal matrix parameter $\bm{Q}$ is matrix Bingham-von Mises-Fisher. \citet{Hoff2009} provides a column-wise strategy for simulating from this conditional posterior distribution.

An approximation to the posterior distribution of the parameters $(c, \bm{Q}, \bm{\Lambda})$ can also be obtained using polar expansion with adaptive HMC as implemented in Stan. To carry out posterior simulation with Stan's adaptive HMC algorithm, we must provide the log posterior density, modulo an additive constant. Applying polar expansion, the log posterior density is
\begin{align}
    \log p(c, \bm{X}, \bm{\Lambda} \mid \bm{Y})  &=\sum_{i > j} y_{i,j} \Phi\left[c + \left(\bm{Q}_X\bm{\Lambda}\bm{Q}_X^\top\right)_{i,j}\right]\nonumber\\  
    &+ \sum_{i > j}(1-y_{i,j})\left\{1- \Phi\left[c + \left(\bm{Q}_X\bm{\Lambda}\bm{Q}_X^\top\right)_{i,j}\right]\right\} \nonumber\\
    &-\frac{c^2}{2 \times 10^2} - \frac{\bm{X}^\top\bm{X}}{2} - \sum_{j=1}^{3}\frac{\lambda_j^2}{2p} + C \label{eigenmodel_density}
\end{align} where $C$ is a constant which does not depend upon the parameters. Given a Markov chain $\{c_t, \bm{X}_t, \bm{\Lambda}_t\}_{t=1}^{T}$ whose stationary distribution has density (\ref{eigenmodel_density}), we approximate the posterior distribution of $(c, \bm{Q}, \bm{\Lambda})$ by $\{c_t, \bm{Q}_{X_t}, \bm{\Lambda}_t\}_{t=1}^{T}.$

\begin{figure}[htbp]
\centerline{\includegraphics[width=6.2in]{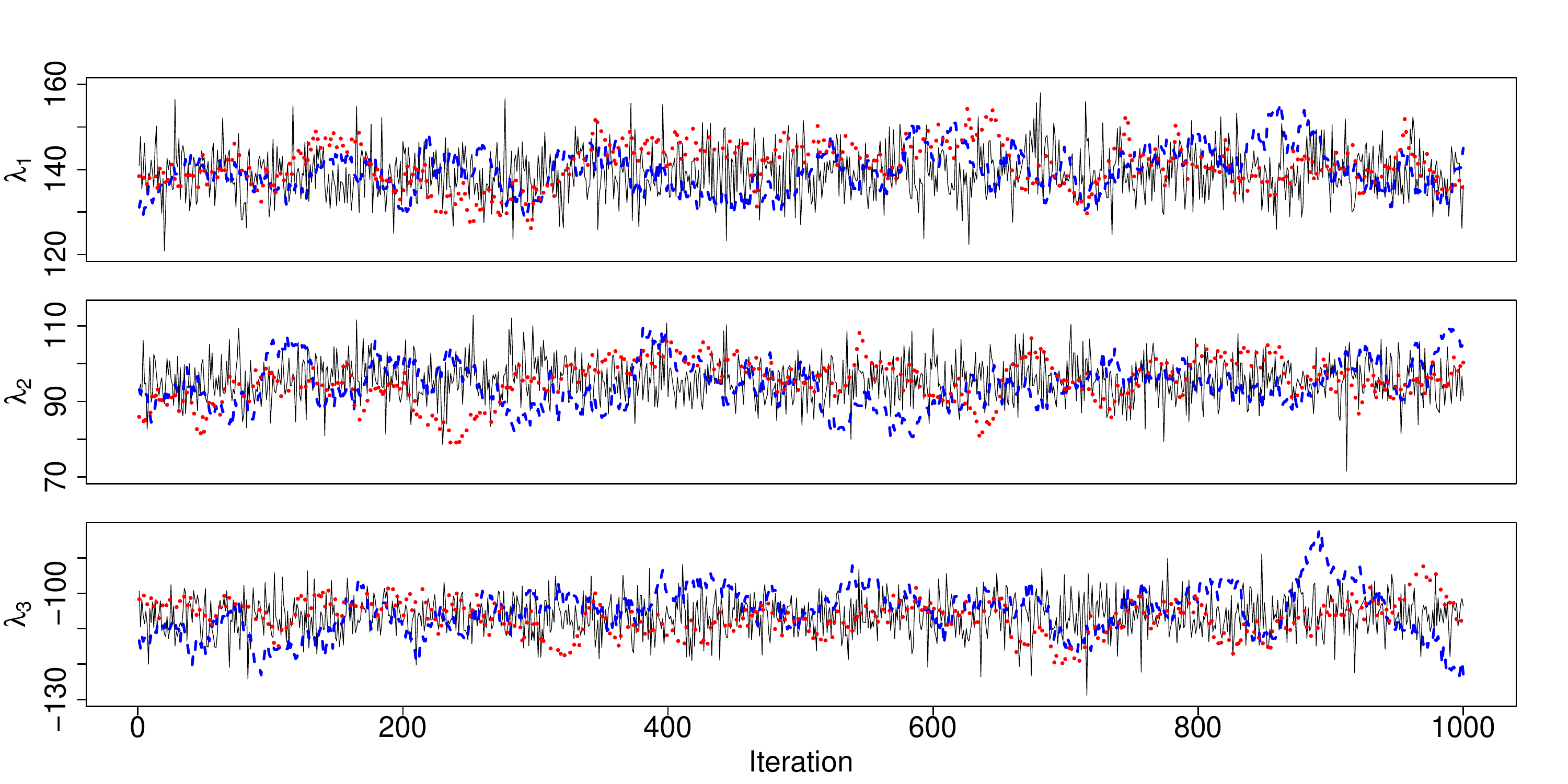}}
\caption{Traceplots for the diagonal elements of $\bm{\Lambda} = \text{diag}(\lambda_1, \lambda_2, \lambda_3)$ based on the three MCMC methods. The solid black lines correspond to polar expansion with adaptive HMC, the dashed blue lines correspond GMC without parallel tempering, and the dotted red lines correspond to the Gibbs sampler of \citet{Hoff2009}.}
\label{fig:eigenmodel_trace}
\end{figure}

Figure \ref{fig:eigenmodel_trace} provides traceplots for the diagonal elements of $\bm{\Lambda} = \text{diag}(\lambda_1, \lambda_2, \lambda_3)$ based on polar expansion with adaptive HMC, GMC without parallel tempering, and the Gibbs sampler of \citet{Hoff2009}. Stan's diagnostics did not give any indication of problems which would lead to poor Monte Carlo estimates. For GMC, we used the tuning parameters given in \citet{Byrne2013}. Even visually, we can tell that the Markov chain produced by polar expansion with adaptive HMC exhibits less autocorrelation than those produced via GMC or the Gibbs sampler. This is confirmed by the calculations in Table \ref{table:ess} which show that the effective sample size per iteration of our approach is an order of magnitude greater than that of the competing methods. Effective sample size per second is the truly relevant quantity to compare, but variability in code quality and random initializations make such comparisons challenging. We remark only that simulating 5000 post warm up Markov chain iterations with our approach took a similar amount of time to the equivalent task with GMC and far less time compared to the Gibbs sampler. %Much of the improvement in effective sample size per iteration in comparison to GMC can be attributed to Stan's adaptive tuning parameter selection functionality. 

Because one can simultaneously permute the columns of $\bm{Q}$ and $\bm{D}$ and change their signs without changing the value of the posterior density, the posterior distribution of the network eigenmodel has multiple symmetric modes. None of the MCMC methods we compare are capable of switching between these symmetric modes. However, this lack of switching does not affect inferences about identifiable parameters. \citet{Byrne2013} combine GMC with parallel tempering and show that the resulting Markov chains do switch between symmetric modes. They also describe how, without parallel tempering, Markov chains produced by GMC can become stuck in a local mode with negligible posterior mass. Markov chains produced by HMC applied to the distribution with the log posterior density (\ref{eigenmodel_density}) are likewise vulnerable to becoming stuck in this mode. However, all the Markov chains in Figure \ref{fig:eigenmodel_trace} have converged to the same mode as in \citet{Byrne2013} and \citet{Hoff2009}.

\begin{table}
\centering
\begin{tabular}{c c c c}
Parameter   & Polar Exp. & GMC & Gibbs  \\
\hline
$\lambda_1$ & 0.835 & 0.031 & 0.030 \\
$\lambda_2$ & 0.886 & 0.038 & 0.030 \\
$\lambda_3$ & 0.683 & 0.033 & 0.036  \\
\end{tabular}
\caption{Effective sample sizes per iteration for the diagonal elements of $\bm{\Lambda}$ calculated using the \texttt{R} package \texttt{mcmcse} \citep{Flegal2017}. The calculations are based on 5000 post warm up Markov chain iterations.}
\label{table:ess}
\end{table}

\subsection{Principal components analysis of functional data} \label{functionalPCA}

We propose a new approach to Bayesian functional principal components analysis which we illustrate in a meteorological time series application.  Principal component analysis linearly transforms a set of high-dimensional, correlated variables into a lower-dimensional set of uncorrelated ``principal component scores," accounting for as much variation in the original data as possible. PCA has become an essential tool for exploratory data analysis and dimension reduction, and has inspired a vast literature of related methodology. When applied to data arising from an underlying curve or surface, however, classical PCA fails to take the functional structure into account and, as a result, can be excessively noisy. Ramsay and Silverman's influential book \citep{Ramsay1997} describes how to adapt PCA to functional data from a penalized optimization perspective. As an alternative, our Bayesian approach to principal components analysis of functional data has a number of potential advantages: functional structure can be incorporated through the prior distribution, smoothing parameters can be estimated rather than chosen via cross-validation, and parameter uncertainty is reflected in posterior distribution. Additionally, our method can easily accommodate certain types of missing data and can be flexibly modified or extended.

\begin{figure}%[H]
\centerline{\includegraphics[width=6.25in]{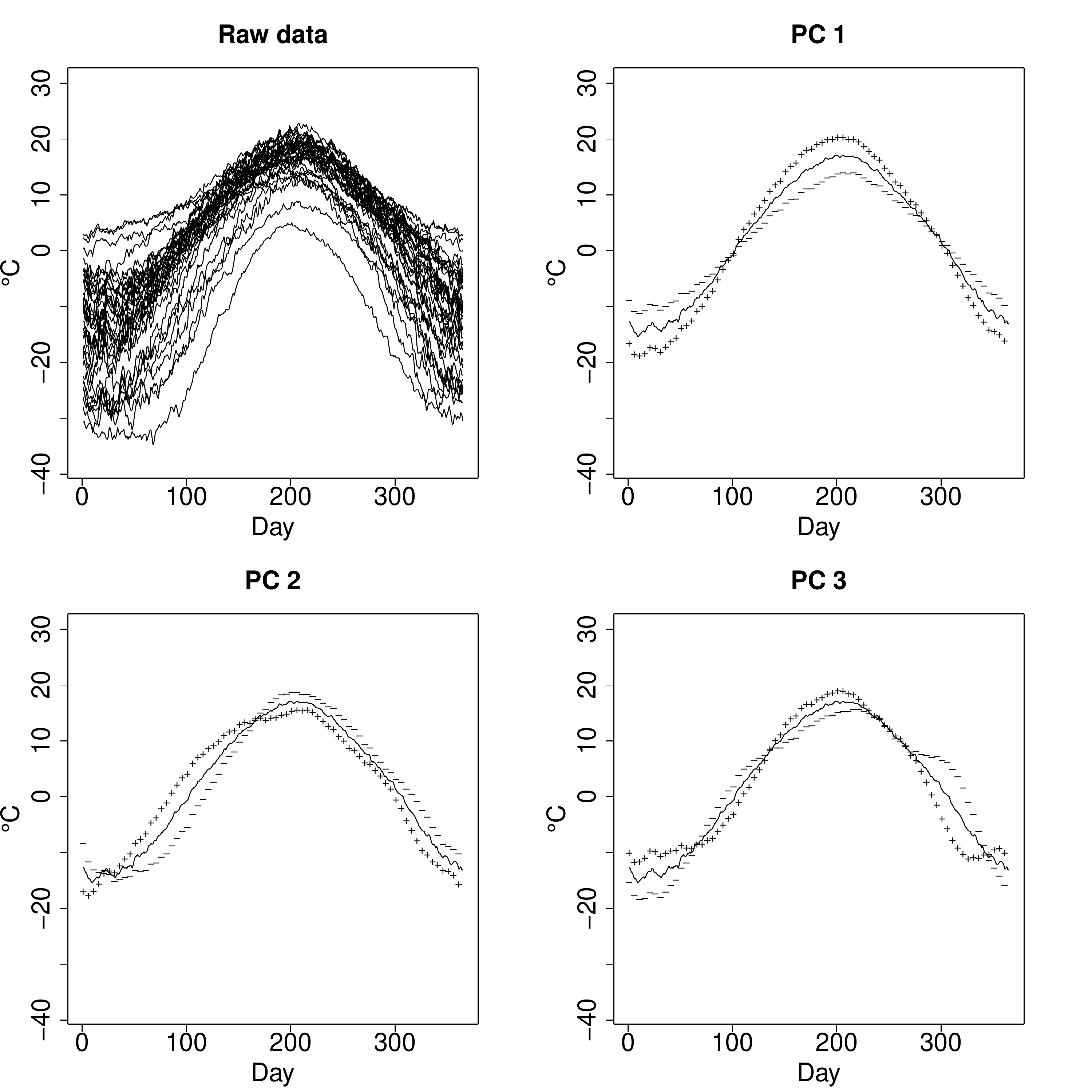}}
\caption{The columns of $\bm{Y}_\text{raw}$ are plotted in the top left panel. The other three panels plot the column means of $\bm{Y}_{\text{raw}}$ plus and minus a suitable multiple of the principal component curves.}
\label{fig:interp}
\end{figure}

We consider the Canadian weather data previously analyzed in \citet{Ramsay1997} and \citet{Suarez2017}. The Canadian weather data set, available in the $\texttt{R}$ \citep{Team2019} package $\texttt{FDA}$ \citep{Ramsay2018}, includes average daily temperatures for 35 weather stations throughout Canada. The raw data matrix $\bm{Y}_{\text{raw}}$ has $n=35$ rows and $p=365$ columns with entry $(i,j)$ recording the average temperature in city $i$ on day $j.$ The columns of $\bm{Y}_{\text{raw}}$ are plotted  in the top left panel of Figure \ref{fig:interp}. Immediately, we see the functional nature of the data, large differences in the average yearly temperature across cities, and a roughly sinusoidal pattern of seasonal variation. Large differences in average yearly temperature are to be expected, given that the data set includes weather stations from Victoria, British Columbia to Inuvik, Northwest Territories. The roughly sinusoidal pattern of seasonal variation is also unsurprising. The aim of our functional principal component analysis is to identify subtler modes of variation present in the data, while taking into account its functional nature. 

We subtract row and column means from $\bm{Y}_{\text{raw}}$ and model the resulting matrix as $\bm{Y} = \bm{U}\bm{D} \bm{V}^\top + \sigma\bm{E}\bm{\Omega}(\varphi)^{1/2}$. The unknown parameters are the orthogonal matrices $\bm{U}\in \mathcal{V}(k,n)$ and $\bm{V} \in \mathcal{V}(k,p),$ the diagonal matrix $\bm{D} = \text{diag}(d_1, ..., d_k)$ with $d_1, ..., d_k > 0,$ the scale parameter $\sigma >0,$ and the correlation parameter $\varphi.$ The entries of the matrix $\bm{E}$ are independent standard normal random variables, and $\bm{\Omega}(\varphi)$ is the correlation matrix of an AR(1) process with parameter $\varphi.$ The low rank matrix $\bm{U}\bm{D} \bm{V}^\top$ is intended to capture long term, seasonal variation in temperature. The rows of $\bm{U}\bm{D}$ contain the principal component scores for each weather station, while the columns of $\bm{V}$ form the corresponding basis of principal component curves. In this analysis, we set $k=3.$ The matrix $\sigma\bm{E}\bm{\Omega}(\varphi)^{1/2}$ is intended to capture short term, day to day variation in temperature. Conditional on $\sigma$ and $\varphi,$ each column of $\sigma\bm{E}\bm{\Omega}(\varphi)^{1/2}$ is an independent AR(1) process.

We assign the parameter $\bm{V}$ a hierarchical prior chosen to reflect the functional nature of the temperature data. Because we intend $\bm{U}\bm{D} \bm{V}^\top$ to capture long term, seasonal variation in temperature, we want the principal component curves in each column of $\bm{V}$ to look like the discretization of a smooth function. This functional structure can be represented by a $\text{MACG}(\bm{K})$ prior when $\bm{K}=(k_{i,j})$ is constructed using, for instance, the squared exponential covariance function \citep{Rasmussen2006} with $k_{i,j} = \exp\left[-(i - j)^2/\rho^2\right].$ The length-scale hyperparameter $\rho$ controls the ``wiggliness" of the principal component curves. We assign $\rho$ an inverse gamma prior, yielding the following hierarchical prior for $\bm{V}:$
\begin{align*}
    \bm{V} \mid \rho &\sim \text{MACG}(\bm{K}) \\ 
    1/\rho &\sim \text{Ga}(\alpha, \beta).
\end{align*} 
When $\bm{V}$ is $\text{MACG}(\bm{K})$ and $p \gg k,$ each column of $\bm{V}$ behaves like a centered Gaussian process (GP) with a squared exponential covariance function and length-scale $\rho.$ %(Justification for this claim is given in the first author's PhD dissertation, which is available upon request.) 
Such a GP is infinitely differentiable, and the expected number of zero crossings in an interval of length $T$ is $T/(2 \pi \rho)$ \citep{Rasmussen2006, Adler1981}. Motivated by the latter observation, we choose the inverse gamma hyperparameters $\alpha$ and $\beta$ so that $\rho$ has a prior mean of $365/(4\pi)$ and prior standard deviation of five. If $\rho$ were fixed at its prior mean, the expected number of zero crossings of the principal component curves would be approximately two. The proposed inverse gamma prior also puts very little prior mass on small values of $\rho.$ Together, these attributes reflect our intention that $\bm{U}\bm{D} \bm{V}^\top$ capture long term, seasonal variation in temperature.

We now specify priors for the remaining parameters. The rows of $\bm{U}$ correspond to locations throughout Canada. We could try to incorporate this spatial structure through the prior distribution, but in this analysis we simply assign $\bm{U}$ a uniform prior. To the correlation parameter $\varphi,$ we assign the arc-sine prior discussed in \citet{Fosdick2012}. The priors for $\sigma^2$ are inverse gamma and truncated normal:
\begin{align*}
    1/\sigma^2 &\sim \text{Ga}\left(\frac{\nu}{2},\frac{\nu}{2} s^2\right) \\ 
    p(d_1, ... , d_k) &\propto \mathbbm{1}\left\{d_1, ... , d_k > 0\right\} \prod_{i=1}^k N(d_i \,;\, 0, \tau^2).
\end{align*} 
We use an empirical Bayes strategy to select the hyperparameters $\nu, s^2,$ and $\tau.$ Let $\widehat{\bm{Y}}$ be the best rank-k approximation to $\bm{Y}$ in the Frobenius norm \cite{Eckart1936}, and let $\widehat{\sigma^2}$ be the sample variance of the entries of the residual matrix $\bm{Y} - \widehat{\bm{Y}}.$ The prior variance of $\sigma^2$ is decreasing as a function of the hyperparameter $\nu,$ which has an interpretation as a prior sample size in a normal model \citep{Hoff2009a}. We let $\nu=1$ and then set $s^2 = 3\,\widehat{\sigma^2}$ so that the prior mode for $\sigma^2$ is $\widehat{\sigma^2}.$ We choose $\tau^2$ so that the prior expectation of $\sum_{i=1}^{k} d_i^2$ is equal to $\Tr(\widehat{\bm{Y}}^\top\widehat{\bm{Y}}).$ 

\begin{figure}[H] %[p]
\centerline{\includegraphics[width=6.25in]{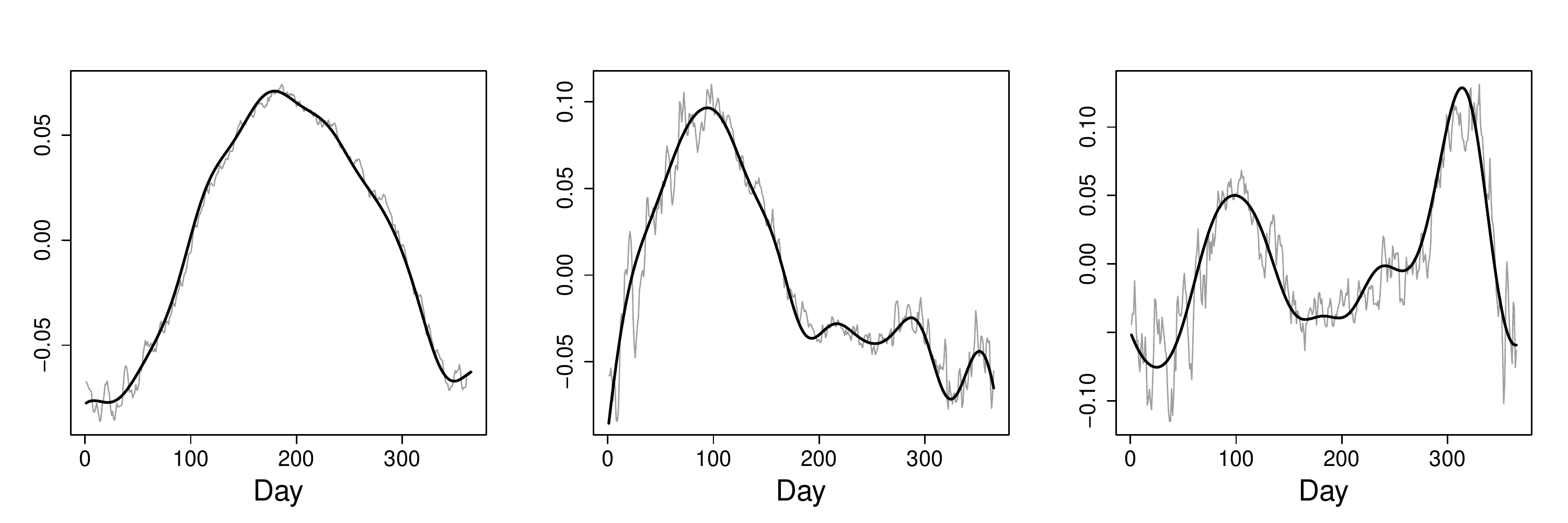}}
\caption{A comparison of our point estimate of $\bm{V}$ to the results of classical PCA. The black lines are our estimated principal component curves, while the gray lines are the corresponding values based on classical PCA.}
\label{fig:MLEcomparison}
\end{figure}

We simulate from the posterior of the proposed functional principal components model using polar expansion with adaptive HMC. Again, Stan's diagnostics did not give any indication of problems which would lead to poor Monte Carlo estimates. As our point estimate of $\bm{V},$ we take the first $k=3$ right singular vectors of the posterior mean of $\bm{U}\bm{D}\bm{V}^\top.$ Figure \ref{fig:MLEcomparison} compares our point estimate to the results of classical PCA. The black lines are our estimated principal component curves, while the gray lines are the corresponding values based on classical PCA. Compared to the results of classical PCA, the principal component curves produced by our method are smoother and less noisy. 

Figure \ref{fig:interp} aids in interpreting the principal component curves. The top left panel is a plot of the raw temperature data. The other three panels plot the column means of $\bm{Y}_{\text{raw}}$ plus and minus a suitable multiple of the principal component curves. The multiple is chosen subjectively for the sake of interpretability. This approach to visualizing principal components analyses of functional data is described in \citet{Ramsay1997}. We see that the first principal component relates to the difference between summer and winter temperatures, with a higher principal component score corresponding to a larger difference. The second principal component relates to a time shift effect. The third principal component is hardest to interpret, but a higher value appears to indicate a later spring and an earlier end to Autumn. 

\begin{figure}[H]
\centerline{\includegraphics[width=6.25in]{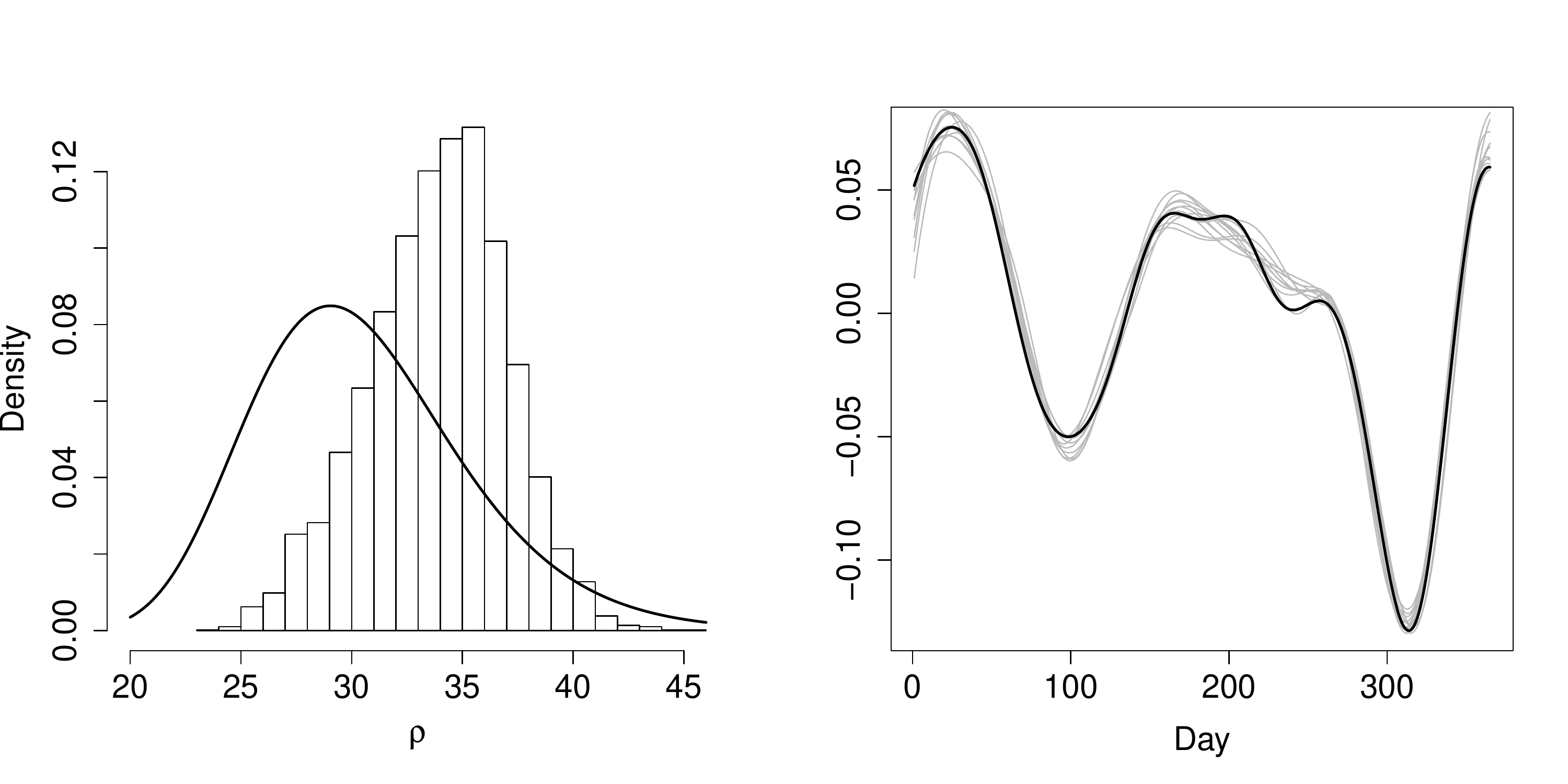}}
\caption{The left hand side compares a histogram estimate of the marginal posterior density of $\rho$ with its prior density. The right hand side shows simulated posterior values of the third principal component curve (in gray) and its point estimate (in black).}
\label{fig:misc}
\end{figure}

The left hand side of Figure \ref{fig:misc} compares a histogram estimate of the marginal posterior density of $\rho$ with its prior density. The marginal posterior distribution of $\rho$ is more concentrated than the prior and has a higher mean, indicating that the proposed method can learn a suitable value for $\rho$ without resorting to cross-validation. The right hand side shows simulated posterior values of the third principal component curve (in gray) and the point estimate (in black). The simulated posterior values, not just the point estimate, are smooth, and their variation about the point estimate reflects the parameter uncertainty remaining.

\section{Discussion} \label{discussion}

Together with modern MCMC software, polar expansion allows for routine and flexible simulation from probability distributions on the Stiefel manifold, including posterior distributions arising from statistical models with orthogonal matrix parameters. The key idea is to transform the constrained simulation problem into an easier unconstrained problem using the polar decomposition and its Jacobian. We described how to apply polar expansion in simulation problems and then considered two applications. In the first, we found that polar expansion with adaptive HMC is an order of magnitude more efficient than competing MCMC approaches in a benchmark protein interaction network application. In the second, we proposed a new approach to Bayesian functional principal components analysis which we illustrated in a meteorological time series application. 

We briefly describe a few directions for future work. Proposition \ref{X_density_prop} tells us we have a great deal of flexibility in our choice of conditional density $f_{S_X \mid Q_X}.$ The choices in Section \ref{MCMC} are motivated by the simplicity of the resulting distribution for $\bm{X}.$ While these choices work well in a wide range of simulation problems, it would be interesting to explore more systematically how the choice of conditional density impacts subsequent MCMC simulation. Thus far, we have made a case for polar expansion based on its practical performance. Recent work on the convergence of HMC may provide tools to analyze polar expansion from a theoretical perspective \citep{Durmus2017, Livingstone2018, Bou-Rabee2017}. Sections \ref{MCMC} and \ref{applications} demonstrate that prior distributions which are the $\bm{Q}_X$-margin of a standard distribution for $\bm{X}$ are tractable and useful. An interesting direction is to study the relationship between the distribution of $\bm{X}$ and its $\bm{Q}_X$-margin.

\section*{Acknowledgements}
This work was supported by the United States Office of Naval Research under Grant N00014-14-1-0245/N00014-16-1-2147.
\bibliographystyle{apalike}
\bibliography{references}

%\appendix

\end{document}